Bistable MAP Kinase Activity: A Plausible Mechanism Contributing to Maintenance of Late Long-Term Potentiation

Paul Smolen, Douglas A. Baxter, and John H. Byrne

Department of Neurobiology and Anatomy
The University of Texas Medical School at Houston
Houston, TX 77225

E-mail: Paul.D.Smolen@uth.tmc.edu

## ABSTRACT

Bistability of MAP kinase (MAPK) activity has been suggested to contribute to several cellular processes, including differentiation and long-term synaptic potentiation. A recent model (48) predicts bistability due to interactions of the kinases and phosphatases in the MAPK pathway, without feedback from MAPK to earlier reactions. Using this model and enzyme concentrations appropriate for neurons, we simulated bistable MAPK activity, but bistability only was present within a relatively narrow range of activity of Raf, the first pathway kinase. Stochastic fluctuations in molecule numbers eliminated bistability for small molecule numbers, such as are expected in the volume of a dendritic spine. However, positive feedback loops have been posited from MAPK up to Raf activation. One proposed loop in which MAPK directly activates Raf was incorporated into the model. We found that such feedback greatly enhanced the robustness of both stable states of MAPK activity to stochastic fluctuations and to parameter variations. Bistability was robust for molecule numbers plausible for a dendritic spine volume. The upper state of MAPK activity was resistant to inhibition of MEK activation for > 1 h, suggesting inhibitor experiments have not sufficed to rule out a role for persistent MAPK activity in LTP maintenance. These simulations suggest that persistent MAPK activity and consequent upregulation of translation may contribute to LTP maintenance and to long-term memory. Experiments using a fluorescent MAPK substrate may further test this hypothesis.

## INTRODUCTION

Considerable evidence supports the hypothesis that long-term potentiation (LTP) of synaptic connections is an essential component of memory formation and maintenance (47). This hypothesis implies memories are at least partly stored as patterns of strengthened synapses. The form of LTP relevant for long-term memory, late LTP (L-LTP), lasts for > 3 h and requires protein synthesis (52, 68, 78). L-LTP can persist for months *in vivo* (1). An outstanding question is, how can some memories be preserved for months or years given turnover of neuronal and synaptic proteins? Synaptic proteins typically have lifetimes of hours to a few days (19).

To preserve strong synapses, some proposed mechanisms hypothesize bistability in the activity of kinases that, when active, strengthen synapses. A stimulus that induced L-LTP could "switch" a bistable kinase into a persistently active state. MAP kinase (MAPK) activity is necessary for upregulation of dendritic protein synthesis following stimulation by NMDA (28). Bistable activity of the MAPK signaling pathway is thus an interesting candidate mechanism for synaptic strength maintenance, because persistent MAPK activation would tend to upregulate local dendritic translation. Raf is the upstream kinase in the MAPK signaling pathway. Active Raf doubly phosphorylates and activates MEK kinase, which doubly phosphorylates and activates MAPK. Persistent activation of these kinases would lead to phosphorylation and activation of newly synthesized MAPK, thereby counteracting loss of MAPK activity due to protein turnover. Bistability of CaM kinase II autophosphorylation and activity (50) or of PKA activity and AMPA receptor phosphorylation (32) have also been hypothesized to participate in maintenance of synaptic strength.

Bistable MAPK activity has been suggested by experiments with various cell types. Prolonged activation of MAPK is necessary and sufficient for neuronal differentiation of PC-12 cells (15). When PC-12 cells and BHK cells were transfected with active Raf, the increased Raf activity generated all-or-none, binary MAPK activation, suggesting bistability of MAPK activity (30). In the BHK cells, transfection of constitutively active MEK activated endogenous MAPK, but did not activate endogenous MEK, suggesting that in this cell type, there is not positive feedback from MAPK through Raf back to MEK. In further experiments (60), sustained (>1 h) activation of MAPK occurred in PC-12 cells following brief applications of nerve growth factor. Dynamics in which a brief stimulus yields a long-lasting state transition characterize bistability.

These empirical observations motivated us to consider further whether bistable MAPK activity may contribute to maintenance of L-LTP and memory. Some previous modeling studies have hypothesized that bistable MAPK activity is important for the consolidation of L-LTP (8, 9, 41). However, these models contain a positive feedback loop from MAPK to Raf (via protein kinase C), and thence to MEK (i.e., MAPK activation leads to Raf activation, which activates MEK). Therefore, these models do not explain the bistable MAPK activity observed in BHK cells (30), which occurs in the absence of feedback from MAPK to MEK. In contrast, one recent model (48, 54) generates bistable MAPK activity without such feedback. This model was suggested to explain bistable MAPK activity in BHK and PC-12 cells (30). Bistability arises instead from interactions between MAPK and its activating kinase (MEK) and inactivating phosphatase. We therefore began by modifying this model, using parameters that describe more specifically L-LTP induction and the dynamics of neuronal MAPK. We examine whether bistable MAPK activity can plausibly occur at synapses, and whether L-LTP induction might trigger a long-lasting transition to a state of high MAPK activity.

We consider excitatory L-LTP at synapses between pyramidal cells, such as Schaffer collateral (SC) synapses or neocortical synapses. The role of MAPK and of local dendritic translation in this form of L-LTP has been characterized. The MAPK isoforms that appear necessary for L-LTP induction are extracellular-regulated kinase (ERK) 1 and 2 (59). Our model represents postsynaptic activation of the ERK pathway. A range of Raf activity was determined within which simulated ERK activity was bistable. Brief activation of Raf could permanently switch ERK activity from a low state to a high state.

Experiments suggest that the activity of <u>bulk</u> cytoplasmic ERK does not remain elevated for more than 30-45 min after L-LTP induction (20, 46, but see 3). Any persistently active ERK would need to be restricted to a small volume, such as a dendritic spine, in order to have avoided detection in those assays. In such small volumes, stochastic fluctuations in molecule numbers may destabilize steady states (10, 66, 71). Biochemical events important for L-LTP induction and maintenance occur within dendritic spines, which have volumes on the order of 0.1 fl (31). A concentration of 1 µM corresponds to only 60 molecules in 0.1 fl. Therefore, we simulated the effects of fluctuations in a spine volume. In initial simulations, bistability was eliminated.

However, the robustness of bistable states to fluctuations can be reinforced by positive feedback loops (12). Studies suggest that in some cell types, positive feedback from ERK to Raf, or from ERK to MEK, is present. In COS-7 cells, doubly phosphorylated ERK can phosphorylate and activate Raf (6), forming a positive feedback loop that could reinforce ERK bistability. Feedback from ERK to MEK mediates bistability in *Xenopus* oocytes (75). Feedback in which ERK activates protein kinase C (PKC) to increase Raf activity has also been suggested (8). As noted above, such feedback from ERK to Raf does not appear to operate in BHK cells (30). Whether such feedback exists in neurons is not yet known. However, we considered whether such feedback could plausibly sustain robust bistability of ERK activity for low molecule numbers, such as are expected in a spine volume. A representation of ERK -> Raf feedback was added to the model. Bistability was greatly strengthened. The bistable region spanned a broad range of values of Raf activity. Both stable states were robust to stochastic fluctuations, with lifetimes of months, for molecule numbers that correspond to ERK and Raf concentrations of ~ 2–3 µM in a spine. Both stable states were preserved for moderate variations of model parameters. Furthermore, persistent ERK activation was stable for MEK inhibition of ~ 1 hr duration, suggesting that previous experiments with the MEK inhibitor PD 98059 have not ruled out persistent ERK activity as important for L-LTP maintenance. Persistent ERK activation and upregulation of local protein synthesis appears to constitute a plausible mechanism for maintaining strengthened synapses.

## METHODS

*Model development*

Neuronal MAPK can be activated by stimulus-induced $Ca^{2+}$ influx acting via CaM kinase I (63), by cAMP (29, 51), or by metabotropic glutamate receptors (76). BDNF also activates ERK (38). All these mechanisms probably contribute to the observed activation of ERK by stimuli that induce L-LTP, such as tetanic or theta-burst electrical stimuli (80) or BDNF exposure (77, 38). Raf activation is the convergence point for these mechanisms of activation of the ERK cascade. In neurons, the most common Raf isoforms are Raf-1 or B-Raf (2, 18). Active Raf phosphorylates MAP kinase kinase (MAPKK or MEK) twice, activating MEK. MEK then phosphorylates ERK twice, activating ERK (Fig. 1A). ERK is phosphorylated on tyrosine and threonine residues. The phosphorylation mechanism is distributive (22). Either residue can be phosphorylated first, yielding the intermediates ERK-PT (phosphothreonine) or ERK-PY (phosphotyrosine). Activation of MEK and ERK is counteracted by phosphatases. MAP kinase phosphatase-3 (MKP3) specifically inactivates the ERK subfamily of MAPKs via distributive dephosphorylation (79). PP2A can also dephosphorylate ERK (37). We therefore chose to denote ERK phosphatase less specifically, as MKP. MEK phosphatase was denoted MKKP.

*Modeling ERK activation*

The authors of (48) considered several model variants describing ERK phosphorylation and dephosphorylation. We used the variant most closely based on experimental data for ERK. This variant was termed the "ERK elementary step level" in (48). It is rather complex, with 13 ordinary differential equations. Elementary reactions (bimolecular association reactions, and unimolecular phosphorylation or dissociation reactions) describe binding of ERK to MEK or MKP, and ERK phosphorylation and dephosphorylation. The reactions are given in Supplementary Table 2 of (48) and in Table 1 of this study. Figure 1B illustrates the reactions of ERK with MEK and with MKP. In Fig. 1B, the asymmetry in the dephosphorylation reactions reflects the empirical mechanism of the common MAP kinase phosphatase MKP3. MKP3 dephosphorylation of ERK-PP follows an ordered mechanism with phosphotyrosine dephosphorylated first (79). In Tables 1-2, for brevity, ERK-PP and MEK-PP are termed EPP and MPP, ERK-PT is termed EPT, and MEK is termed M.

*Modeling MEK activation*

Experimentally, MEK activation kinetics have not been as well analyzed. We used an elementary-step description of MEK phosphorylation and dephosphorylation. However, instead of copying the kinetic scheme specific to ERK from Table 1, we used the somewhat simpler set of reactions in Supplementary Table 1 of (48). Although the authors denote that model variant as a description of the "MAPK cycle", it serves equally well as a plausible description of MEK phosphorylation. Because the relative abundances of different isoforms of MEK, as well as detailed mechanisms, are not well characterized, this intermediate level of detail seems reasonable. The MEK reactions are given in Table 2 of this study. The MEK reactions (Table 2) and the ERK reactions (Table 1) are coupled via the phosphorylation of ERK by MEK. Standard values for all reaction rate constants are in Tables 1 and 2. $Raf_{ACT}$ denotes the amount of active Raf capable of phosphorylating MEK.

Table 3 presents the 26 model differential equations. The reaction rates on the right-hand sides of these equations are given in Tables 1 and 2. Table 3 also gives assumed values for the concentrations of total MEK (both free and bound, both inactive and active), total ERK, total MKKP, and total MKP. The model does not include protein synthesis or degradation. Table 3 also shows conservation conditions for each enzyme.

*Addition of positive feedback from ERK-PP to Raf activation*

We developed an extended model variant with a positive feedback loop in which active ERK phosphorylated and activated Raf (dashed line, Fig. 1A). Recent data (6) suggest several possible sites for ERK phosphorylation (S239, S296, S301). Whether all these sites are phosphorylated is not known. We chose to represent an intermediate degree of feedback by assuming ERK-PP phosphorylates two sites to activate Raf. In the absence of feedback, Raf activity is a parameter, but with ERK -> Raf phosphorylation, the states of Raf are dynamic variables. The two reactions in the first line of Table 4 describe the successive phosphorylations of Raf by active ERK (ERK-PP). As in Tables 1-2, the simplest plausible kinetics are used, with a second-order rate constant $k_{f,Raf}$ describing phosphorylation and a first-order rate constant describing dephosphorylation. Three differential equations result for unphosphorylated, singly phosphorylated, and doubly phosphorylated Raf (Table 4). There is a conservation condition for total Raf (parameter $Raf_{TOT}$, third line of Table 4). Parameter values describing Raf dynamics are given in the fourth line of Table 4.

To allow for the relatively simple representation of Raf dynamics in Table 4, the following assumptions were made: binding of MEK to Raf does not affect Raf phosphorylation, Raf activation by stimuli and phosphorylation by ERK are independent processes, a Raf molecule is activated equally by stimuli and by ERK phosphorylation, Raf is not further activated by both events together, and only activated Raf binds MEK. Current data do not necessitate the more complex model that would be required by relaxing these assumptions. The assumptions allow us to express the activity of Raf, $Raf_{ACT}$, as a sum of two terms (last line of Table 4). The first is a parameter describing the amount of Raf activated by an applied stimulus ($Raf_{STIM}$). The second term is the amount of doubly phosphorylated Raf multiplied by the ratio of <u>unstimulated</u> Raf to total Raf. Because only active Raf is assumed to bind MEK, $Raf_{ACT}$ is used directly in the kinetic expressions of Table 1.

In all simulations illustrated in Figs. 2-8, the standard parameter values from Tables 1-4 are used, except as noted in the text and figure legends.

*Simulating activation of Raf, MKKP, and MKP*

Following a theta-burst stimulus that induces hippocampal LTP, BDNF secretion is upregulated for ~ 12 min (4). BDNF activates ERK (11, 38). LTP-inducing stimuli may also activate ERK through pathways dependent on $Ca^{2+}$ influx, but activation by BDNF is likely to have a slower time course. Therefore, to simulate ERK activation due to L-LTP induction, we increased the parameter that denotes the amount of stimulus-activated Raf for ~ 12 min. The increase was a simple square wave, returning to basal level after the stimulus. For the simulations of Figs. 2-3, the parameter is $Raf_{ACT}$, the amount of active Raf. For the simulations of Figs. 4-7 with the extended model variant, $Raf_{ACT}$ is a dynamic variable (Table 4, last line). Therefore, the parameter increased was the amount of Raf specifically activated by the stimulus, $Raf_{STIM}$ (Table 4, last line).

In Figs. 2, 4, 5, and 6, temporary activations of ERK phosphatase (MKP) were also simulated to illustrate how stimuli might destabilize elevated ERK activity. Amplitudes and durations of MKP increase are given in the text or figure legends. After the increase, the amount of MKP capable of interacting with ERK is returned to its basal value (the standard parameter $MKP_{TOT}$) by brief rapid removal of free MKP.

*Simulation of stochastic fluctuations in molecule numbers*

Fluctuations were simulated with the Gillespie algorithm. This algorithm takes variable time steps, and during each time step, exactly one reaction occurs. Which type of reaction occurs is determined randomly, with the probability of each reaction type proportional to its deterministic rate expression as given in Tables 1, 2, and 4. For further algorithm details see (26, 27). In stochastic simulations, the molecule copy numbers are scaled by using a volume factor that multiplies or divides concentration and rate parameters. In order to multiply the average molecule numbers by $f_{stoch}$, total enzyme concentrations ($MEK_{tot}$ etc.) are multiplied by $f_{stoch}$. Second-order rate constants describing are divided by $f_{stoch}$ (*e.g.*, $k_{11f}$ and $k_{13f}$ in Table 1, $k_{f,Raf}$ in Table 4). First-order rate constants remain unchanged (*e.g.*, $k_{11b}$ and $k_{12f}$ in Table 1, $k_{b,Raf}$ in Table 4). Values of $f_{stoch}$ are given in the figure legends.

*Numerical methods*

For deterministic simulations (Figs. 2, 4) the forward Euler method was used for integration of differential equations, with a time step of 15 msec. Simulations verified further time step reductions did not significantly improve accuracy. To further verify accuracy, the simulation of Fig. 4A was repeated using the second-order Runge-Kutta integration method (57). No significant differences were observed in any of the variables. Prior to any stimulus, variable values were determined by equilibration. First, all variables were initialized to zero except for free nonphosphorylated Raf, MEK, ERK, MKKP, and MKP, which were set equal to the total amounts of the respective enzyme ($Raf_{TOT}$, $MEK_{TOT}$, $ERK_{TOT}$, $MKKP_{TOT}$, $MKP_{TOT}$). Then, all variables were equilibrated for at least four simulated days, establishing steady-state levels of active ERK, Raf, and the other variables. We verified that longer equilibrations did not significantly alter these levels. The model was programmed in Java and simulated on Pentium 3 microcomputers. Programs are available upon request.

Bifurcation analysis examined how steady states of ERK activity ([ERK-PP]) depend on the amount of active Raf. The software package MATCONT was used (available at http://www.matcont.ugent.be/, see also http://portal.acm.org/citation.cfm?id=779362). For the bifurcation diagram of Fig. 2B, the bifurcation parameter is Raf activity ($Raf_{ACT}$). For the diagram of Fig. 4B, Raf activity is a dynamic variable, therefore the bifurcation parameter is the amount of Raf activated by an applied stimulus ($Raf_{STIM}$ in the last line of Table 4). During bifurcation analysis, to obtain convergence with MATCONT, it was necessary to make use of the conservation conditions for each enzyme (Tables 3 and 4). Each conservation relation was used to eliminate the differential equation for the free (unbound) form of that enzyme, by expressing free enzyme as (total enzyme) – (all bound forms). This procedure is commonly necessary for numerical bifurcation analysis of kinetic systems.

## RESULTS

*Simulation of bistability and state transitions*

Figure 2A illustrates bistability and state transitions simulated with the model of Fig. 1, without positive feedback from ERK to Raf. At *t* = 30 min, Raf is activated. The parameter $Raf_{ACT}$, denoting total active Raf (free and bound to MEK), is abruptly increased for 15 min from 120 to 600 nM. As a result, MEK and subsequently ERK are doubly phosphorylated and activated (the levels of [MEK-PP] and [ERK-PP] increase). When $Raf_{ACT}$ is returned to basal levels, active MEK and ERK remain in the stable upper state. At *t* = 2 h, the activity of the ERK phosphatase MKP is increased. The level of free MKP available for interaction with ERK is increased by 140 nM, thus total [MKP] increases from 220 to 360 nM. This increase lasts for 40 min, after which $MKP_{TOT}$ is returned to 220 nM by rapid removal of free MKP. The upper state of MEK and ERK activity is destabilized by this increase in $MKP_{TOT}$.

The bifurcation diagram of Fig. 2B illustrates bistability of ERK activity as a function of active Raf concentration. Parameters are as in Fig. 2A except $Raf_{ACT}$, which is the bifurcation parameter. In a relatively narrow range of $Raf_{ACT}$, between 106 and 139 nM, two stable steady states of ERK-PP exist (upper and lower branches of curve). They are separated by a middle unstable steady state (middle branch). The bistability region is bounded by limit points (LP). At each LP the unstable state merges with one of the stable states, eliminating both. The bifurcation analysis software MATCONT (see Methods) indicates that two neutral saddle points exist on the unstable state, and denotes them HB. In the bistable region, large system perturbations, such as brief increases or decreases in Raf$_{ACT}$, can switch the system between steady states.

For our model, the values of three kinetic rate constants were modified from (48). These constants are k$_{17f}$, k$_{19f}$, and k$_{20f}$ in Table 1. Original values (48) are in parentheses. The changes to k$_{19f}$ and k$_{20f}$ are minor and lie within experimental uncertainty. These changes slightly increased the range of Raf$_{ACT}$ that supports bistability. To further enhance the bistability range, k$_{17f}$ was increased to 0.02 s$^{-1}$. This increase appears consistent with data. It decreases the Michaelis constant (K$_m$) for the reaction ERK-PT -> ERK-PP to 75 nM. Experimental estimates of this K$_m$ are dependent on cell type. The authors of (48) used data from *Xenopus* oocyte p42 MAPK (22) to estimate K$_m$ = 300 nM, whereas the authors of (8) used K$_m$ = 46 nM as estimated for *E. coli* p42 MAPK (33). Multiple ERK isoforms are present in neurons, thus the physiological K$_m$ is uncertain, and 75 nM appears reasonable. As a check of these modified rate constants, we repeated the simulation of Supplemental Figure 1 of (48), which compares ERK phosphorylation time courses with data. The simulated time courses of threonine-phosphorylated ERK (all subspecies) and of tyrosine-phosphorylated ERK were not significantly altered, and their fit to the data was not degraded.

For Figs. 2A-B, total concentrations of each enzyme (MEK$_{TOT}$, ERK$_{TOT}$, MKP$_{TOT}$, MKKP$_{TOT}$, Table 3) were similar to estimates of bulk cytoplasmic ERK and MEK concentrations (8). To try to increase the range of bistability, we varied these concentrations within relatively broad ranges (from ~ 50% to 200% of the values in Table 3). In all cases the Raf$_{ACT}$ bistability range remained relatively narrow, never significantly broader than in Fig. 2B.

*Bistability is disrupted by stochastic fluctuations for small molecule numbers*

To further assess whether the mechanism of (48) could plausibly sustain bistable ERK activity within a small volume like that of a dendritic spine, stochastic simulations were carried out using the Gillespie algorithm (see Methods). Binding events and reaction events occur stochastically, changing molecule copy numbers. To simulate fluctuations, it is first necessary to estimate the average numbers of relevant molecules present in the system being simulated. Typical spine volumes are ~ (500 nm)$^3$ = 0.1 fl (31). One μM corresponds to 60 molecules in 0.1 fl. For PC-12 cells, ([ERK], [MEK], [Raf]) have been reported in μM as (0.26, 0.68, 0.5) (61). Some neuronal models have assumed values of 0.2 – 0.4 μM for all three kinases (8, 9). However, data suggest concentrations of ERK, MEK, and Raf in spines may be considerably above the values for bulk cytoplasm (see Discussion). We therefore allowed average molecule numbers for ERK, MEK, and Raf to vary up to ~200, which corresponds to 3-4 μM in a spine.

For molecule numbers in this range, we were unable to simulate robust bistability. Figure 3A illustrates a typical simulation, with the molecule number scaling factor f$_{stoch}$ set to 0.3 nM$^{-1}$ (details in Methods). Simulated time scales were long (~ 10 days) to model dynamics relevant for long-term memory maintenance. At *t* = 90 h, Raf was activated for 15 min. During this interval only, the total number of active Raf molecules (both free and bound to MEK) was increased from its basal value of 36 up to 180. Except for this brief imposed increase in Raf, the total numbers of Raf, MEK, and ERK molecules are conserved. Molecule numbers of MEK and ERK (all forms) were respectively 198 and 210. ERK activity appeared to settle into an upper state. However, the upper state was not stable to stochastic fluctuations. At *t* = 230 h, a spontaneous fluctuation destabilized the upper state of ERK-PP, and ERK activity returned to the lower state. Similar results were obtained for other choices of model parameters, and in most cases the lifetime of the upper state was less.

The lifetime decreases rapidly for smaller molecule numbers (lower $f_{stoch}$). For molecule numbers corresponding to concentrations of 2 µM or less in a spine, the upper state lifetime never exceeded 2 h in 20 simulations with all other model parameters as in Fig. 3A. In contrast, if molecule numbers are scaled up, stability of the upper state is readily achieved. For example, with total MEK, ERK, and Raf numbers doubled from Fig. 3A (*i.e.* $f_{stoch}$ doubled to 0.6 nM$^{-1}$), the upper state remained stable for at least 300 h in 20 out of 20 simulations.

Although these latter molecule numbers appear too high to be relevant for conditions in a dendritic spine, bistability of ERK activity may be important for processes such as cell differentiation or oncogenesis (see Discussion). Therefore, for these higher molecule numbers, we analyzed the stability of the upper and lower states of ERK activity to variations in model parameters. The control simulations for the upper and lower states had parameters as in Fig. 3A except molecule numbers were increased by setting the scaling factor $f_{stoch}$ to 0.6 nM$^{-1}$. In a series of simulations, the 47 model parameters in Tables 1-3 were successively varied from the control values by +30% and -30%. For each of these 94 parameter variations, two stochastic simulations were carried out. In one, ERK-PP and MEK-PP were initialized in the lower steady state, and in the other, ERK-PP and MEK-PP were initialized in the upper state. Therefore, 95 pairs of simulations were carried out (two pairs for each parameter, and one control pair with standard parameter values). During each simulation, variables were first equilibrated for 100 h, and the average number of doubly phosphorylated ERK molecules over the next 100 h was determined.

The result is illustrated in Fig. 3B. Each plotted point corresponds to one of the 95 pairs of simulations. The simulation initialized in the lower steady state of ERK-PP has its time-average number of ERK-PP molecules plotted on the x-axis, and the simulation initialized in the upper steady state has its time-average ERK-PP plotted on the y-axis. The scatter plot shows that most parameter variations do not significantly change the high and low steady states of ERK-PP. Most of the 95 points are clustered near the grey point corresponding to control values of all parameters. For each point in this cluster, the low value of x illustrates that a simulation stabilized in the low state, and the high value of y illustrates that the simulation stabilized in the high state. In Fig. 3B, there is a smaller cluster of points near the origin. These points correspond to simulations in which the high steady state of ERK-PP was destabilized by the parameter increase or decrease. For these simulations, initial conditions near the high state evolved to the low state, corresponding to a low value on the y axis. There are 10 such points. There are also 7 points in a cluster in the upper right part of the graph (to the right of 200 on the x-axis) corresponding to simulations in which only the lower steady state of ERK-PP was destabilized.

*Positive feedback allows preservation of bistability for small molecule numbers*

The robustness of bistable states to stochastic noise is expected to be enhanced by the presence of multiple, reinforcing positive feedback loops (12). The authors of (6) suggest ERK-PP can doubly phosphorylate and activate Raf, forming such a positive feedback loop. A representation of this feedback was added to the model. The equations and parameters describing this feedback are given in Table 4 (Methods). Figure 4A illustrates bistability with this extended model. At $t = 30$ min, Raf is stimulated for 17 min. The amount of Raf activated by external stimulus ($Raf_{STIM}$ in Table 4) is increased from a basal value of 75 nM to 600 nM. The Raf time course in Fig. 4A is total active Raf available for interaction with MEK. This variable, $Raf_{ACT}$, is a sum of $Raf_{STIM}$ and of doubly phosphorylated Raf (Methods and Table 4). When Raf activation ceases, ERK-PP remains in the upper state. At $t = 2$ h, the ERK phosphatase MKP is activated. Free MKP is increased by 140 nM for 30 min, after which $MKP_{TOT}$ is returned to its basal value (220 nM) by rapid removal of free MKP. The upper state of ERK activity is thereby destabilized.

The bifurcation diagram of Fig. 4B illustrates bistability of ERK activity ([ERK-PP]). Parameters are as in Fig. 4A except that the amount of Raf activated by an applied stimulus is varied. Thus, $Raf_{STIM}$ is the bifurcation parameter. The positive feedback from ERK to Raf allows a large bistability range. For $Raf_{STIM}$ between 0 and 115 nM, two stable states of ERK-PP exist. MATCONT failed to indicate the limit points or any saddle points. However, we verified by simulations that the stable steady states were accurately determined. Brief, large system perturbations can switch the system between steady states. However, the upper steady state remains stable even if $Raf_{STIM}$ is reduced to zero, because ERK phosphorylation maintains Raf activity (with $Raf_{STIM} = 0$, $Raf_{ACT}$ in Table 4 equals [Raf-PP]). To switch the system to the lower state, phosphatase activation is required.

Bistability of the extended model with stochastic fluctuations was simulated. The additional positive feedback allowed the upper and lower steady states of ERK activity to be stable at much lower molecule numbers. Fig. 5A illustrates a long simulation in which ERK activation was repeatedly induced and removed by applied stimuli. Both states of ERK activity were stable in the absence of stimuli. At $t = 50$ h, the number of activated Raf molecules was increased from its low baseline of 15 up to 120. The increase lasted for 12 min. This stimulus induced MEK and ERK activity, and both activities transited to a stable, fluctuating upper state. At $t = 220$ h, the number of ERK phosphatase (MKP) molecules was increased by 27, by abruptly increasing free MKP. The increase lasted for 30 min, after which 27 molecules of free MKP were removed. This brief increase in MKP activity sufficed to induce a state transition, with MEK and ERK activity returning to the lower state. At $t = 280$ h and $t = 340$ h, the brief Raf activation and brief MKP activation were respectively repeated. These state transitions are reproducible, as is the stability of both states. In 20 repetitions of the simulation of Fig. 5A, differing only in the initialization of the random number generator used in the Gillespie algorithm, the same dynamics were obtained. In Fig. 5A, the "averaged ERK-PP" time course refers to the average of ERK-PP over these 20 simulations.

Figures 5B-C illustrate dynamics of additional variables during this simulation. Figure 5B illustrates activated MEK (MEK-PP), active Raf, and free Raf (not bound to MEK). The time course of MEK-PP is like that of ERK-PP. Because ERK-PP activates Raf, the time course of active Raf is also like ERK-PP. Free Raf increases slightly when MEK-PP increases, because MEK-PP is not a substrate for Raf. At $t = 170$ h, Raf was briefly activated with the same stimulus parameters as for $t = 50$ h. Because ERK and MEK are already active, the only visible effect at $t = 170$ h is the brief spike in active Raf (asterisk). Figure 5C shows that free MEK drops to near zero in the upper state of ERK and MEK activity. This drop is because almost all MEK is bound to either Raf-PP or MKKP. Similar dynamics are observed for free ERK. In the upper state most ERK is bound to MKP3 phosphatase or to MEK-PP.

*Bistability is robust to variations in model parameters*

Figure 6A illustrates that the upper and lower steady states of the simulation of Fig. 5 are robust to moderate variations in model parameter values. The 50 model parameters in Tables 1-4 were successively varied by +30% and -30%. The effect of these variations on both the high and low steady states of active, doubly phosphorylated ERK was examined. The plot was constructed in the same manner as Fig. 3B. Therefore, 101 pairs of simulations were carried out including the control pair with all parameters standard. For each pair, the simulation initialized in the lower steady state of ERK-PP has its time-average number of ERK-PP molecules plotted on the x-axis, and the simulation initialized in the upper steady state has its average ERK-PP plotted on the y-axis. The plot shows that almost all parameter variations do not significantly change the high and low steady states of ERK-PP. Of the 101 points, all except 4 are clustered near the grey control point. Here, simulations initialized in the low steady state of ERK-PP remained in the low state, and simulations initialized in the upper steady state remained there. For two points to the upper right, the lower steady state was lost, and for two points near the origin, the upper steady state was lost. The two points near the origin correspond respectively to 30% decreases in $ERK_{TOT}$ and in $MEK_{TOT}$. The two points to the upper right correspond respectively to 30% decreases in $MKP_{TOT}$ and in $MKKP_{TOT}$ (ERK and MEK phosphatase levels). Thus the model dynamics are particularly sensitive to variations in the amounts of ERK, MEK, MKKP, and MKP.

Figure 6B repeats the robustness simulations of Fig. 6A with a key difference. For each of the simulations with parameter variation, all 50 parameters were varied. For each parameter in each simulation, the variation was a Gaussian random variable. The mean variation was zero and the standard deviation was set to 10% of the control parameter value. The Box-Mueller algorithm was used to compute these variations (57). The resulting scatter plot demonstrates that even with simultaneous variations of all parameters, both the upper and lower steady states of ERK-PP were preserved in most simulations. Most of the 101 points are still clustered near the grey control point. Nine points are clustered to the right of ERK-PP = 80 on the x-axis, corresponding to destabilization of the lower steady state, and seven points are clustered near the origin, corresponding to loss of the upper steady state.

Stability of bistable ERK activity to fluctuations requires that the positive feedback in which ERK phosphorylates Raf be of sufficient strength. In the first equation in Table 4, the rate of Raf phosphorylation by ERK is proportional to the rate constant $k_{f,Raf}$. Therefore, feedback strength is governed by $k_{f,Raf}$. Data does not currently allow the strength of any ERK -> Raf positive feedback to be determined in neurons. However, it is important to examine whether the simulated feedback strength is physiologically plausible. To this end, we carried out a simulation of the time course of Raf phosphorylation when the amount of active ERK was switched from 0 to 80% of total ERK. Parameters were as in Fig. 5 with three exceptions. ERK-PP was clamped at 0 prior to t = 1 h and at 80% of total ERK thereafter. Stimulus-induced Raf activation separate from ERK ($Raf_{STIM}$) was set to 0.0. To isolate the time course of Raf phosphorylation from the process of Raf binding to MEK, the Raf -> MEK association constants ($k_{1f}$, $k_{3f}$) were set to 0.0. Figure 7A illustrates that the amplitude and time course of Raf activation appears physiologically plausible. Over 30-40 min, non-phosphorylated Raf declines from 100% of total Raf to ~30%. Active, doubly phosphorylated Raf increases from 0 to ~40% of $Raf_{TOT}$. Thus, Raf phosphorylation by ERK is substantial but not near completion, and the duration required to reach a fluctuating plateau appears reasonable.

We also examined the robustness of the bistability in Fig. 5 to changes in the strength of positive feedback. $k_{f,Raf}$ was varied over a range extending from 50% up to 300% of its control value in Fig. 5A (steps of 10% were used). For each value of $k_{f,Raf}$, the system was initialized in the lower state of ERK-PP and MEK-PP. No perturbation was applied until $t = 300$ h, at which time the number of activated Raf molecules was increased from its baseline of 15 up to 120, for 20 min. The system transited into the upper state, with high ERK-PP and MEK-PP. The stability of the upper state was monitored for another 300 h. The simulation was repeated 5 times for each $k_{f,Raf}$ value. Both the upper and lower states were stable for a rather broad range of $k_{f,Raf}$ values. Bistability was preserved when $k_{f,Raf}$ was at least 70% of control and at most 240% of control. Thus, bistability is relatively robust to variations in the strength of positive feedback.

*Elevated ERK activity is resistant to inhibition of MEK for > 1 h*

Inhibition of MEK activation by PD 98059 for ~ 1 h blocks L-LTP induction but does not reverse established L-LTP (20, 35). This result could mean prolonged ERK activity is not in fact necessary to maintain L-LTP or LTM. However, it is also possible that ERK activity is necessary and that the extent or duration of MEK inhibition did not suffice to destabilize elevated ERK activity. We examined the robustness of the upper state of ERK-PP (Fig. 5A) to temporary inhibition of MEK activation. PD 98059 is a noncompetitive inhibitor of MEK. PD 98059 binds to inactive, nonphosphorylated MEK. PD 98059 does not appear to bind to active MEK (5, 21) and we assumed PD 98059 also does not bind singly phosphorylated MEK (current data do not appear to test this assumption). We simulated inhibition of MEK by adding an inactive molecular species, MEK-PD, to the model. MEK-PD forms by association of free nonphosphorylated MEK with PD 98059, and MEK-PD dissociates to give free MEK. We denote the concentration of PD 98059 as [PD]. Binding of PD 98059 to MEK is half-maximal at a PD 98059 concentration of 4 ± 2 μM (5). In the model we assign the mean value, 4 μM, to the dissociation constant of PD 98059 from MEK. The association and dissociation rates of MEK-PD, $R_{f,inh}$ and $R_{b,inh}$, are given by the following mass action expressions

$$R_{f,inh} = k_{f,inh}[MEK][PD], \quad R_{b,inh} = k_{b,inh}[MEK-PD]$$

$k_{b,inh}$ and $k_{f,inh}$ were chosen to have values of 10.0 $sec^{-1}$ and 2.5 $\mu M^{-1} sec^{-1}$, respectively. The dissociation constant of PD 98059 from MEK is the ratio of $k_{b,inh}$ to $k_{f,inh}$, 4 μM. Simulation results were relatively insensitive to scaling the values of $k_{f,inh}$ and $k_{b,inh}$ as long as their ratio was not changed. The stochastic simulation of Fig. 5A was redone with the formation and dissociation of MEK-PD included. Initially, [PD] = 0, and ERK-PP and MEK-PP were initialized in the upper state. After 1 h in the stable upper state, [PD] was increased to 50 μM for a duration of 4,000 sec, during which time MEK-PD could form. After this interval, [PD] returned to zero.

Figure 7B illustrates that this inhibition did not destabilize the upper state of elevated ERK activity. Only a brief, minor drop in MEK-PP and ERK-PP was observed. Each time course is an average over 20 simulations that differ only in the initial random number generator seed. ERK was resistant to inhibition in all 20 simulations. We also verified that this degree of MEK inhibition blocked the induction of persistent ERK activity. If the simulation of Fig. 5A was repeated with PD 98059 applied during Raf activation, only slight transient activation of ERK resulted (not shown).

Stronger inhibition of MEK did destabilize ERK activation. If the simulation of Fig. 7B was repeated with $k_{f,inh}$ = 1.0 $\mu M^{-1}sec^{-1}$, corresponding to a tighter MEK – inhibitor dissociation constant of 1.0 $\mu M$, then a 1 h inhibitor application drove ERK activity to the lower state.

## DISCUSSION

We developed a model to aid in examining whether the mechanism proposed by (48) and (54) is a plausible generator of bistable ERK activity in a small compartment with a volume similar to a dendritic spine. We first verified bistability in deterministic simulations (Fig. 2A), with enzyme concentrations in the ERK signaling pathway set to values within the range of estimates for neuronal cytoplasm. However, the region of bistability was not large. To sustain bistability, basal (unstimulated) Raf activity had to lie within a relatively narrow range (Fig. 2B). Using this mechanism, the authors of (43) and (48) also obtained a relatively narrow range of bistability.

A challenge for models suggesting biochemical bistability contributes to maintenance of L-LTP is to preserve stable steady states despite stochastic fluctuations in molecule numbers. Fluctuations are significant in small biochemical volumes relevant for L-LTP, such as a dendritic spine. Indeed, for a previous model of bistable MAPK activity (8), bistability was overwhelmed by stochastic fluctuations in a spine volume (10). In our first model variant, for numbers of ERK signaling pathway molecules that might be expected in a spine volume, bistability was eliminated by fluctuations (Fig. 3A). However, for larger molecule numbers, bistability was preserved and was relatively robust against moderate (30%) variations in model parameters (Fig. 3B). Such molecule numbers are relevant for nuclei, where ERK bistability has been suggested to play roles in regulation of gene expression, cell differentiation (23), or oncogenesis (42).

These simulations failed to support the hypothesis that persistent ERK activation contributes to synaptic strength maintenance. However, we considered that additional feedback might be present *in vivo* and might reinforce the robustness of bistability. The authors of (6) present evidence for positive feedback in which ERK activates Raf. We incorporated this feedback into the model. The parameter range with bistability was broadened (Fig. 4). Indeed, bistability remained when stimulus-induced Raf activation was removed (Fig. 4B, $Raf_{STIM}$ = 0)*.*

Figure 5A illustrates that with positive feedback, bistable ERK activity was robust to stochastic fluctuations at low molecule numbers. State transitions were induced by stimuli of plausible duration. Activation of Raf for 12 min caused ERK activity to transit to the upper state. During experimental L-LTP induction, theta-burst stimuli increase BDNF secretion for about 12 min (4). BDNF application induces L-LTP (77), and activates ERK (11). Also in Fig. 5A, ERK activity transited to the lower state following activation of ERK phosphatase for 30 min.

In Fig. 5A, both states of ERK activity are stable with the following total enzyme molecule numbers: $Raf_{TOT}$ = 125, $MEK_{TOT}$ = 144, $ERK_{TOT}$ = 140, $MKKP_{TOT}$ = 44, $MKP_{TOT}$ = 44. In the volume of a spine (~ 0.1 fl), these numbers correspond to concentrations of ~ 1–3 $\mu M$. Such concentrations appear plausible for spines. Immunohistochemistry has been used to assess Raf in dendritic spines from cortex and hippocampus (49). Strong staining of numerous spines was observed, and was clearly enhanced over the neighboring dendrite. Western blotting has determined that MEK and ERK2 are localized in synaptosomal fractions and in postsynaptic density (PSD) fractions from cortex and hippocampus (69). Strong colocalization of the PSD protein PSD-95 with ERK in hippocampal neurons has been seen (36). Therefore, the simulation of Figs. 5A-C suggests that a biochemically plausible ERK -> Raf positive feedback loop could sustain bistable ERK activity, and long-term activation of ERK, within a spine volume. Figures 6A-B illustrate that even with these small molecule numbers, the upper and lower states of ERK activity are generally preserved for moderate variations in model parameters.

If an LTP – inducing stimulus persistently activated ERK, how might this translate into prolonged synaptic strengthening? L-LTP induction has been shown to lead to ERK phosphorylation of the initiation factor eIF4E (7). MAPK activity is necessary for upregulation of dendritic protein synthesis (28, 39). ERK acts through the mTOR pathway to upregulate translation (70). During induction of hippocampal LTP, polyribosomes move into spines (55). Therefore, sustained ERK activation in a spine or in a dendrite would be expected to enhance local protein synthesis, plausibly increasing synaptic weights. L-LTP – inducing electrical stimuli also set a synaptic tag that "marks" stimulated synapses (24, 25). The tag involves covalent modifications that allow a synapse to "capture" plasticity factors (proteins or mRNAs). Synaptic ERK activity could constitute a tag component. ERK-dependent upregulation of local translation may allow the synapse to use mRNA "plasticity factors" to increase synaptic strength.

Data concerning the duration of ERK activation after induction of L-LTP have not yet been obtained for small compartments such as dendritic spines. For bulk neuronal cytoplasm, the majority of observations (20, 46, but see 3) suggest an activation duration of only ~30 min after tetanus. At the level of dendrites or spines, quantitation of the strength and duration of ERK activity might be feasible using changes in fluorescence resonance energy transfer (FRET) in a designed substrate. A genetically encoded fluorescent indicator ERK substrate, termed Erkus, has recently been developed (62). It is plausible that Erkus could be transfected and expressed in hippocampal slice preparations or in cultured neurons. Following L-LTP induction, the time course of ERK activation within a restricted region, such as a section of a dendrite, could be analyzed with such a probe, thereby providing an experimental test of our model. However, it is not yet known whether this method can resolve ERK activation within a region as small as a dendritic spine. Fluorescent Raf or MEK substrates, if developed, might similarly be used.

Bistability in Figs. 5-7 relies on the positive feedback from ERK to Raf suggested by (6). An earlier study with HEK 293 cells also identified a positive feedback loop in which MEK acts through ERK to enhance Raf phosphorylation and activity (81). However, other authors (17, 34) suggest that in contrast, ERK inhibits Raf. Therefore, the type and role (if any) of ERK -> Raf feedback in neurons needs to be clarified by further experiments. However, we chose activation of Raf by ERK only as one possible representation of positive feedback in the ERK signaling cascade. To sustain bistability, the feedback needed to be of sufficient strength. Although data determining the strength of any ERK -> Raf feedback are not available for neurons, we found that the simulated feedback strength yielded a biochemically reasonable time course for Raf phosphorylation following ERK activation (Fig. 7A). We also found that the simulated bistability was relatively robust to changes in the strength of positive feedback.

Other positive feedback loops might connect ERK activity to Raf or MEK activity *in vivo*. One example is a MAPK -> PKC -> Raf feedback loop, which plays a central role in generation of bistable MAPK activity in several recent models of L-LTP consolidation (8, 9, 41). The individual components of this loop have been demonstrated experimentally. PKC stimulates MAPK via activation of Raf (74), MAPK activates phospholipase A2 (PLA2) (44), and arachidonic acid produced by PLA2 stimulates PKC (53). This type of indirect ERK -> Raf positive feedback could help to sustain bistability in a small volume such as a spine. MAP kinase cascades may also display oscillatory behavior if negative feedback is present. These dynamics have been theoretically analyzed for several types of feedback (14, 58).

Inhibition of the MAP kinase pathway by PD 98059 does not reverse L-LTP once established (20, 35), suggesting persistent ERK activity might not in fact be necessary to sustain late LTP. However, PD 98059 does not inhibit ERK directly, but rather inhibits activation of MEK. Incomplete inhibition of MEK activation, or inhibition limited in duration, may not suffice to destabilize a steady state of activated ERK. The dissociation constant of PD 98059 from MEK is ~ 5 $\mu$M. Simulated inhibition of MEK by PD 98059 for > 1 h did not destabilize elevated ERK activity (Fig. 7B). However, if the simulation of Fig. 7B was repeated with a tighter MEK – inhibitor dissociation constant of 1.0 $\mu$M, then 1 h of inhibition did destabilize ERK activation. The affinity for MEK of another inhibitor, U0126, is ~ 0.5 $\mu$M (21). Our model therefore suggests that application of U0126 might be observed to disrupt L-LTP maintenance by eliminating persistent ERK activity.

Previous modeling studies suggesting a role for MAPK bistability in L-LTP maintenance (8, 9, 41) did not consider implications of inhibitor experiments. It would be of interest to examine whether sustained MAPK activation in those models is robust to MEK inhibition similar to that of Fig. 7B. In another recent model (67), synaptic weight increases were long-lasting due to a large time constant for synaptic weight decay (~3 months). However, no mechanism for generating this time constant was specified.

Recent investigations (56, 64) have implicated persistent activation of an atypical protein kinase C isoform, protein kinase M$\zeta$ (PKM$\zeta$), as necessary for the maintenance of late LTP and at least some forms of LTM. Inhibition of PKM$\zeta$ eliminates late LTP and LTM. Introducing active PKM$\zeta$ into postsynaptic neurons was found to be sufficient to induce early LTP (45), but PKM$\zeta$ activity has not yet been shown to be sufficient on its own to maintain late LTP and LTM. ERK and PKM$\zeta$ activation could both be necessary to maintain late LTP. A finding in future experiments that postsynaptic injection of active PKM$\zeta$ suffices for late LTP would indicate persistent ERK activity could only be important if ERK was downstream of PKM$\zeta$ or in a positive feedback loop with PKM$\zeta$. Persistent ERK activity could contribute to long-term synaptic strengthening is by increasing the synthesis rate of PKM$\zeta$ in the vicinity of the synapse. MAPK activity supports translation of PKM$\zeta$ mRNA (40).

Further study of the biochemical mechanisms underlying long-term synaptic strengthening, such as sustained local translation and enhanced activation of upstream kinases, is necessary to elucidate mechanisms of memory storage. Recent experimental and theoretical work suggests periodic reactivation of memories is also an essential component of the mechanism of memory maintenance (16, 65, 73). Understanding the interface of synaptic biochemical mechanisms with periodic reactivation of memory will be a key to understanding the preservation of memories for up to a lifetime in animals including humans.

This work was supported by National Institutes of Health Grant P01 NS38310 (to J. H. Byrne).

Smolen et al.78. **Young JZ, Nguyen PV.** Homosynaptic and heterosynaptic inhibition of synaptic tagging and capture of long-term potentiation by previous synaptic activity. *J. Neurosci.* 25: 7221-7231, 2005.

79. **Zhao Y, Zhang ZY.** The mechanism of dephosphorylation of extracellular signal-regulated kinase 2 by mitogen-activated protein kinase phosphatase 3. *J. Biol. Chem.* 276: 32382–32391, 2001.

80. **Zhao M, Adams JP, Dudek SM.** Pattern-dependent role of NMDA receptors in action potential generation: consequences on extracellular signal-regulated kinase activation. *J. Neurosci.* 25: 7032-7039, 2005.

81. **Zimmermann S, Rommel C, Ziogas A, Lovric J, Moelling K, Radziwill G.** MEK1 mediates a positive feedback on Raf-1 activity independently of Ras and Src. *Oncogene* 15: 1503-1511, 1997.
## FIGURE LEGENDS

**Figure 1.** Schematic of the model for activation of the ERK kinase cascade. (A). Schematic of the overall ERK cascade. Active Raf doubly phosphorylates MEK and MAP kinase kinase phosphatase (MKKP) deactivates MEK. The model assumes distributive phosphorylation of MEK by Raf, and distributive dephosphorylation of both MEK-PP and ERK-PP by MKKP and MKP respectively. Dashed curved lines denote putative positive feedback in which ERK activates Raf. This feedback is used in some subsequent simulations (Figs. 4-7). (B). Phosphorylation of ERK by doubly phosphorylated MEK (MEK-PP), and dephosphorylation by MAP kinase phosphatase (MKP). MEK-PP can phosphorylate ERK on either a threonine or tyrosine residue, generating two singly phosphorylated ERK species, ERK-PY and ERK-PT. MEK-PP phosphorylates these intermediates to generate ERK-PP. MKP dephosphorylates the tyrosine residue in ERK-PP, generating ERK-PY. MKP can dephosphorylate either ERK-PY or ERK-PT to regenerate free ERK.

**Figure 2.** Bistable ERK activity with the model of Fig. 1, without positive feedback from ERK to Raf. (A). Simulation of bistability. At $t = 0.5$ h, the concentration of active Raf was briefly increased from a basal value of 120 to 600 nM, for 15 min. Active ERK and MEK transited to a stable upper state. Most ERK was active (ERK-PP ~ 400 nM, $ERK_{TOT}$ = 700 nM) as was most MEK. At $t = 2$ h, the total MKP available to dephosphorylate ERK was increased from its basal value of 220 to 360 nM. The increase lasted for 40 min. [ERK-PP] and [MEK-PP] returned to the lower state. (B). A bifurcation diagram illustrating bistability of ERK activity in the model of Fig. 1 without positive feedback. If [$Raf_{ACT}$] is between 106 and 139 nM, three steady states of ERK activity ([ERK-PP]) exist. The lower and upper states are stable to small perturbations of system variables or parameters. The middle state is unstable to perturbations. Limit points (LP) denote edges of the bistable region. HB denotes neutral saddle points.

**Figure 3.** Stochastic simulations without positive feedback. (A). Time courses of free unphosphorylated ERK and of active doubly phosphorylated ERK. At $t = 90$ h, Raf was activated for 15 min. During this interval only, the total number of active Raf molecules was increased

from its basal value of 36, to 180. ERK activity was elevated. However, at $t = 230$ h, a spontaneous fluctuation destabilized the upper state of ERK-PP. The scaling factor $f_{stoch}$ is 0.3 nM$^{-1}$. (B). Robustness of upper and lower states of ERK activity. With control parameter values both states are stable because $f_{stoch}$, and therefore average molecule numbers, are doubled from (A). The parameters in Tables 1-3 were successively varied by +30% and -30%. For each variation, two simulations were carried out. One simulation started with ERK-PP and MEK-PP in the upper state and the other started with ERK-PP and MEK-PP in the lower state. During each simulation, model variables were equilibrated for 100 h to determine if parameter variation destabilized the initial state. The average ERK-PP over the next 100 h was then used for the plot. Each point corresponds to a distinct parameter variation and the two corresponding simulations (including a control point with no variation, grey square). The y coordinate corresponds to the simulation initialized in the upper state, and the x coordinate to initialization in the lower state.

**Figure 4.** Bistability in the extended model with ERK -> Raf positive feedback. (A). Simulation illustrating state transitions. Initially [ERK-PP] is low and the parameter describing the level of Raf activated by an external stimulus, Raf$_{STIM}$, is low (75 nM). At $t = 30$ min, Raf$_{STIM}$ is increased to 600 nM for 17 min. ERK and MEK activities transited to a stable upper state. At $t = 2$ h, total MKP was increased from its basal value of 220 nM to 360 nM, for 30 min. [ERK-PP] and [MEK-PP] returned to the lower state. For simulations with the extended model (Figs. 5-7), total MEK concentration (MEK$_{TOT}$ in Table 3) is set to 720 nM. (B). Bifurcation diagram. ERK activity is bistable over a broad range of stimulated Raf (Raf activated by an external stimulus, parameter Raf$_{STIM}$ in Table 4).

**Figure 5.** Bistability in the extended model with stochastic fluctuations. (A). Time courses of free unphosphorylated ERK and of active, doubly phosphorylated ERK (ERK-PP). At $t = 50$ h the number of activated Raf molecules was increased from its low baseline of 15 up to 120, for 12 min. ERK-PP and MEK-PP transited to a stable, fluctuating upper state. At $t = 220$ h the number of ERK phosphatase (MKP) molecules was increased by 27. The increase lasted for 30 min. MEK and ERK activity returned to the lower state. At $t = 280$ h and 340 h the brief Raf activation and the brief MKP activation were respectively repeated. The red time course is ERK-PP averaged over 20 simulations. For simulations with the extended model (Figs. 5-8), the scaling factor $f_{stoch} = 0.2$ nM$^{-1}$. (B). Free Raf (not bound to MEK), MEK-PP, and active Raf. Active Raf is calculated by the expression in the last line of Table 4. (C). Time courses of Raf-PP, free MEK, and the average of free MEK over 20 simulations. MEK$_{TOT}$ = 720 nM.

**Figure 6.** Robustness of bistability to parameter variations. (A). Robustness of upper and lower states of ERK activity. The scatter plot is constructed like Fig. 3B, except that 12 simulations were added with variations of the parameters Raf$_{TOT}$, $k_{f,Raf}$, and $k_{b,Raf}$ in Table 4. Most points are very near the control point (grey square). (B). Robustness of states to Gaussian parameter perturbations. The plot was constructed as in (A), except that for each simulation other than the control, <u>every</u> model parameter was perturbed by a Gaussian random variable. The mean of this variable was zero and the standard deviation was 10% of the unperturbed value.

**Figure 7.** Strength of ERK activation by Raf, and robustness of ERK activation to MEK inhibition. (A). The model is initialized with active ERK (ERK-PP) clamped to 0 and with no activation of Raf. At t = 1 h, ERK-PP is clamped to 80% of total ERK. Unphosphorylated Raf then declines by approximately 70% over 30 min, and settles to a noisy plateau. Active, doubly

phosphorylated Raf increases and plateaus over a similar time period, and then fluctuates between 40 and 50 molecules. (B). Robustness to MEK inhibition. The model is initialized in the upper state of (ERK-PP, MEK-PP). Parameters are as in Fig. 5A with $Raf_{STIM} = 15$. At $t = 1$ h, application of 50 µM PD 98059 is simulated as described in the text. The inhibition lasts for 4,000 sec. ERK-PP and MEK-PP remain persistently activated. These variables, and free Raf, decrease slightly during inhibition and then recover. The time courses are averages over 20 simulations.

**TABLE 1. Description of the ERK cycle (Fig. 1B) at the elementary step level**

| Elementary reaction | Reaction rate equation | Rate constants |
|---|---|---|
| E + MPP ↔ E-MPP | $v_{11} = k_{11f} \cdot [E] \cdot [MPP]$ <br> $- k_{11b} \cdot [E\text{-}MPP]$ | $k_{11f} = 0.005 \text{ nM}^{-1}\text{s}^{-1}$ <br> $k_{11b} = 1 \text{ s}^{-1}$ |
| E-MPP → EPY + MPP | $v_{12} = k_{12f} \cdot [E\text{-}MPP]$ | $k_{12f} = 1.08 \text{ s}^{-1}$ |
| EPY + MPP ↔ EPY-MPP | $v_{13} = k_{13f} \cdot [EPY] \cdot [MPP]$ <br> $- k_{13b} \cdot [EPY\text{-}MPP]$ | $k_{13f} = 0.025 \text{ nM}^{-1}\text{s}^{-1}$ <br> $k_{13b} = 1 \text{ s}^{-1}$ |
| EPY-MPP → EPP + MPP | $v_{14} = k_{14f} \cdot [EPY\text{-}MPP]$ | $k_{14f} = 0.007 \text{ s}^{-1}$ |
| E + MPP ↔ E-MPP* | $v_{15} = k_{15f} \cdot [E] \cdot [MPP]$ <br> $- k_{15b} \cdot [E\text{-}MPP^*]$ | $k_{15f} = 0.05 \text{ nM}^{-1}\text{s}^{-1}$ <br> $k_{15b} = 1 \text{ s}^{-1}$ |
| E-MPP* → EPT + MPP | $v_{16} = k_{16f} \cdot [E\text{-}MPP^*]$ | $k_{16f} = 0.008 \text{ s}^{-1}$ |
| EPT + MPP ↔ EPT-MPP | $v_{17} = k_{17f} \cdot [EPT] \cdot [MPP]$ <br> $- k_{17b} \cdot [EPT\text{-}MPP]$ | $k_{17f} = 0.02\ (0.005) \text{ nM}^{-1}\text{s}^{-1}$ <br> $k_{17b} = 1 \text{ s}^{-1}$ |
| EPT-MPP → EPP + MPP | $v_{18} = k_{18f} \cdot [EPT\text{-}MPP]$ | $k_{18f} = 0.45 \text{ s}^{-1}$ |
| EPP + MKP ↔ EPP-MKP | $v_{19} = k_{19f} \cdot [EPP] \cdot [MKP]$ <br> $- k_{19b} \cdot [EPP\text{-}MKP]$ | $k_{19f} = 0.05\ (0.045) \text{ nM}^{-1}\text{s}^{-1}$ <br> $k_{19b} = 1 \text{ s}^{-1}$ |
| EPP-MKP → EPT-MKP | $v_{20} = k_{20f} \cdot [EPP\text{-}MKP]$ | $k_{20f} = 0.085\ (0.092) \text{ s}^{-1}$ |
| EPT-MKP ↔ EPT + MKP | $v_{21} = k_{21f} \cdot [EPT\text{-}MKP]$ <br> $- k_{21b} \cdot [EPT] \cdot [MKP]$ | $k_{21f} = 1 \text{ s}^{-1}$ <br> $k_{21b} = 0.01 \text{ nM}^{-1}\text{s}^{-1}$ |
| EPT + MKP ↔ EPT-MKP* | $v_{22} = k_{22f} \cdot [EPT] \cdot [MKP]$ <br> $- k_{22b} \cdot [EPT\text{-}MKP^*]$ | $k_{22f} = 0.01 \text{ nM}^{-1}\text{s}^{-1}$ <br> $k_{22b} = 1 \text{ s}^{-1}$ |
| EPT-MKP* → E-MKP* | $v_{23} = k_{23f} \cdot [EPT\text{-}MKP^*]$ | $k_{23f} = 0.5 \text{ nM}^{-1}\text{s}^{-1}$ |
| E-MKP* ↔ E + MKP | $v_{24} = k_{24f} \cdot [E\text{-}MKP^*]$ <br> $- k_{24b} \cdot [E] \cdot [MKP]$ | $k_{24f} = 0.086 \text{ s}^{-1}$ <br> $k_{24b} = 0.0011 \text{ nM}^{-1}\text{s}^{-1}$ |
| EPY + MKP ↔ EPY-MKP | $v_{25} = k_{25f} \cdot [EPY] \cdot [MKP]$ <br> $- k_{25b} \cdot [EPY\text{-}MKP]$ | $k_{25f} = 0.01 \text{ nM}^{-1}\text{s}^{-1}$ <br> $k_{25b} = 1 \text{ s}^{-1}$ |
| EPY-MKP → E-MKP | $v_{26} = k_{26f} \cdot [EPY\text{-}MKP]$ | $k_{26f} = 0.47 \text{ s}^{-1}$ |
| E-MKP ↔ E + MKP | $v_{27} = k_{27f} \cdot [E\text{-}MKP]$ <br> $- k_{27b}[E] \cdot [MKP]$ | $k_{27f} = 0.14 \text{ s}^{-1}$ <br> $k_{27b} = 0.0018 \text{ nM}^{-1}\text{s}^{-1}$ |

**TABLE 2. Description of the MEK cycle (Fig. 1A) at the elementary step level**

| Elementary reaction | Reaction rate equation | Rate constants |
|---|---|---|
| $M + Raf_{ACT} \leftrightarrow M\text{-}Raf_{ACT}$ | $v_1 = k_{1f} \cdot [M] \cdot [Raf_{ACT}]$ $- k_{1b} \cdot [Raf_{ACT}]$ | $k_{1f} = 0.02\ nM^{-1}s^{-1}$ $k_{1b} = 1\ s^{-1}$ |
| $M\text{-}Raf_{ACT} \rightarrow MP + Raf_{ACT}$ | $v_2 = k_{2f} \cdot [M\text{-}Raf_{ACT}]$ | $k_{2f} = 0.01\ s^{-1}$ |
| $MP + Raf_{ACT} \leftrightarrow MP\text{-}Raf_{ACT}$ | $v_3 = k_{3f} \cdot [MP] \cdot [Raf_{ACT}]$ $- k_{3b} \cdot [MP\text{-}Raf_{ACT}]$ | $k_{3f} = 0.032\ nM^{-1}s^{-1}$ $k_{3b} = 1\ s^{-1}$ |
| $MP\text{-}Raf_{ACT} \rightarrow MPP + Raf_{ACT}$ | $v_4 = k_{4f} \cdot [MP\text{-}Raf_{ACT}]$ | $k_{4f} = 15\ s^{-1}$ |
| $MPP + MKKP \leftrightarrow MPP\text{-}MKKP$ | $v_5 = k_{5f} \cdot [MPP] \cdot [MKKP]$ $- k_{5b} \cdot [MPP\text{-}MKKP]$ | $k_{5f} = 0.045\ nM^{-1}s^{-1}$ $k_{5b} = 1\ s^{-1}$ |
| $MPP\text{-}MKKP \rightarrow MP\text{-}MKKP$ | $v_6 = k_{6f} \cdot [MPP\text{-}MKKP]$ | $k_{6f} = 0.092\ s^{-1}$ |
| $MP\text{-}MKKP \leftrightarrow MP + MKKP$ | $v_7 = k_{7f} \cdot [MP\text{-}MKKP]$ $- k_{7b} \cdot [MP] \cdot [MKKP]$ | $k_{7f} = 1.0\ s^{-1}$ $k_{7b} = 0.01\ nM^{-1}s^{-1}$ |
| $MP + MKKP \leftrightarrow MP\text{-}MKKP^*$ | $v_8 = k_{8f} \cdot [MP] \cdot [MKKP]$ $- k_{8b} \cdot [MP\text{-}MKKP^*]$ | $k_{8f} = 0.01\ nM^{-1}s^{-1}$ $k_{8b} = 1\ s^{-1}$ |
| $MP\text{-}MKKP^* \rightarrow M\text{-}MKKP$ | $v_9 = k_{9f} \cdot [MP\text{-}MKKP^*]$ | $k_{9f} = 0.5\ s^{-1}$ |
| $M\text{-}MKKP \leftrightarrow M + MKKP$ | $v_{10} = k_{10f} \cdot [M\text{-}MKKP]$ | $k_{10f} = 0.086\ s^{-1}$ $k_{10b} = 0.0011\ nM^{-1}s^{-1}$ |

\* In Tables 1 and 2, species with asterisks denote alternative molecule forms. For example, in Table 1, E-MPP\* reacts to form EPT + MPP, whereas E-MPP (no \*) reacts to form EPY + MPP.

# TABLE 3. Equations of the model, and conserved enzyme totals

$$\frac{d[M]}{dt} = v_{10} - v_1, \quad \frac{d[M\text{-}Raf_{ACT}]}{dt} = v_1 - v_2, \quad \frac{d[MP]}{dt} = v_2 - v_3 + v_7 - v_8,$$

$$\frac{d[MP\text{-}Raf_{ACT}]}{dt} = v_3 - v_4, \quad \frac{d[Raf_{ACT}]}{dt} = v_2 - v_1 + v_4 - v_3,$$

$$\frac{d[MPP]}{dt} = v_4 - v_5 - v_{11} + v_{12} - v_{13} + v_{14} - v_{15} + v_{16} - v_{17} + v_{18},$$

$$\frac{d[MPP\text{-}MKKP]}{dt} = v_5 - v_6, \quad \frac{d[MP\text{-}MKKP]}{dt} = v_6 - v_7, \quad \frac{d[MP\text{-}MKKP^*]}{dt} = v_8 - v_9,$$

$$\frac{d[M\text{-}MKKP]}{dt} = v_9 - v_7, \quad \frac{d[MKKP]}{dt} = -v_5 + v_7 - v_8 + v_{10}, \quad \frac{d[E]}{dt} = -v_{11} - v_{15} + v_{24} + v_{27},$$

$$\frac{d[E\text{-}MPP]}{dt} = v_{11} - v_{12}, \quad \frac{d[EYP]}{dt} = v_{12} - v_{13} - v_{25}, \quad \frac{d[EYP\text{-}MPP]}{dt} = v_{13} - v_{14},$$

$$\frac{d[EPP]}{dt} = v_{14} + v_{18} - v_{19}, \quad \frac{d[E\text{-}MPP^*]}{dt} = v_{15} - v_{16}, \quad \frac{d[ETP]}{dt} = v_{16} - v_{17} + v_{21} - v_{22},$$

$$\frac{d[ETP\text{-}MPP]}{dt} = v_{17} - v_{18}, \quad \frac{d[EPP\text{-}MKP]}{dt} = v_{19} - v_{20}, \quad \frac{d[EPT\text{-}MKP]}{dt} = v_{20} - v_{21},$$

$$\frac{d[EPT\text{-}MKP^*]}{dt} = v_{22} - v_{23}, \quad \frac{d[E\text{-}MKP^*]}{dt} = v_{23} - v_{24}, \quad \frac{d[EPY\text{-}MKP]}{dt} = v_{25} - v_{26},$$

$$\frac{d[E\text{-}MKP]}{dt} = v_{26} - v_{27}, \quad \frac{d[MKP]}{dt} = -v_{19} + v_{21} - v_{22} + v_{24} - v_{25} + v_{27},$$

$ERK_{TOT} = [E] + [E\text{-}MPP] + [EYP] + [EYP\text{-}MPP] + [EPP] + [E\text{-}MPP^*] + [ETP] + [ETP\text{-}MPP]$
$\quad + [EPP\text{-}MKP] + [EPT\text{-}MKP] + [EPT\text{-}MKP^*] + [E\text{-}MKP^*] + [EPY\text{-}MKP] + [E\text{-}MKP]$

$MEK_{TOT} = [M] + [M\text{-}Raf_{ACT}] + [MP] + [MP\text{-}Raf_{ACT}] + [MPP] + [MPP\text{-}MKKP] + [MP\text{-}MKKP]$
$\quad + [MP\text{-}MKKP^*] + [M\text{-}MKKP] + [E\text{-}MPP] + [EYP\text{-}MPP] + [E\text{-}MPP^*] + [ETP\text{-}MPP]$

$MKKP_{TOT} = [MKKP] + [MPP\text{-}MKKP] + [MP\text{-}MKKP] + [MP\text{-}MKKP^*] + [M\text{-}MKKP]$

$MKP_{TOT} = [MKP] + [EPP\text{-}MKP] + [EPT\text{-}MKP] + [EPT\text{-}MKP^*] + [E\text{-}MKP^*]$
$\quad + [EPY\text{-}MKP] + [E\text{-}MKP]$

$MEK_{TOT} = 660$ nM; $\quad ERK_{TOT} = 700$ nM; $\quad MKKP_{TOT} = 220$ nM; $\quad MKP_{TOT} = 220$ nM

## TABLE 4. Equations and parameters describing ERK -> Raf positive feedback

$v_{28} = k_{f,Raf}[\text{Raf}][\text{EPP}] - k_{b,Raf}[\text{RafP}], \quad v_{29} = k_{f,Raf}[\text{RafP}][\text{EPP}] - k_{b,Raf}[\text{RafPP}],$

$\dfrac{d[\text{Raf}]}{dt} = -v_{28}, \quad \dfrac{d[\text{RafP}]}{dt} = v_{28} - v_{29}, \quad \dfrac{d[\text{RafPP}]}{dt} = v_{29},$

$\text{Raf}_{TOT} = [\text{Raf}] + [\text{RafP}] + [\text{RafPP}],$

$\text{Raf}_{TOT} = 660 \text{ nM}, \quad k_{f,Raf} = 3.33 \times 10^{-6} \text{ nM}^{-1}\text{s}^{-1}, \quad k_{b,Raf} = 0.00167 \text{ s}^{-1}$

$[\text{Raf}_{ACT}] = \text{Raf}_{STIM} + [\text{RafPP}] \cdot \dfrac{(\text{Raf}_{TOT} - \text{Raf}_{STIM})}{\text{Raf}_{TOT}},$

Smolen et al.
Figure 1

A.

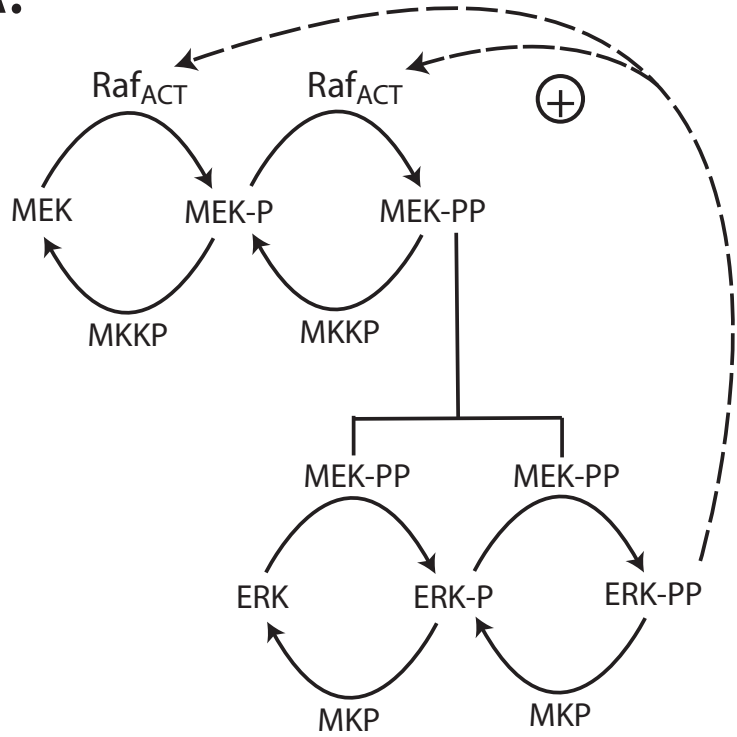

B.

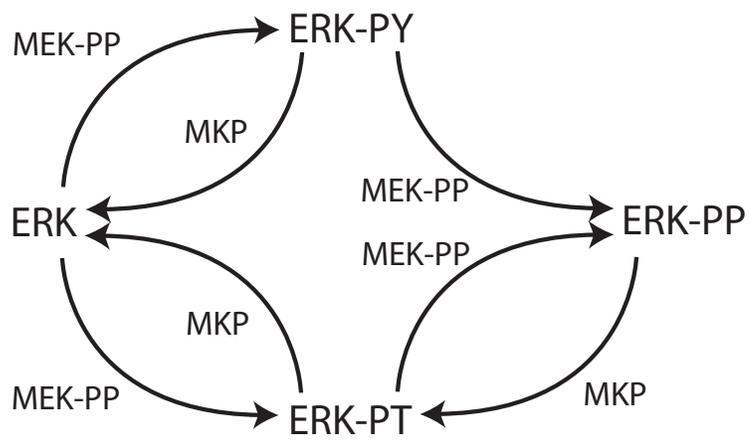

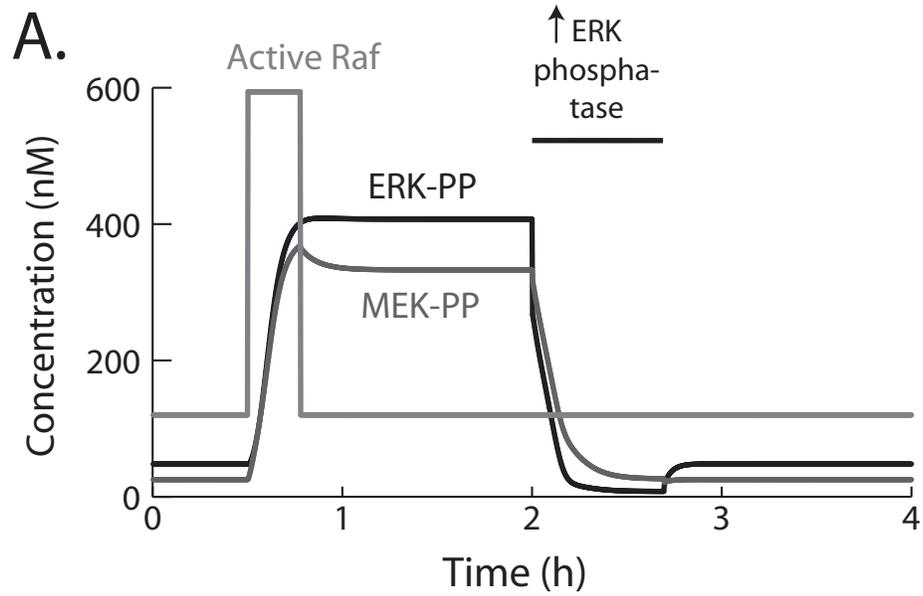

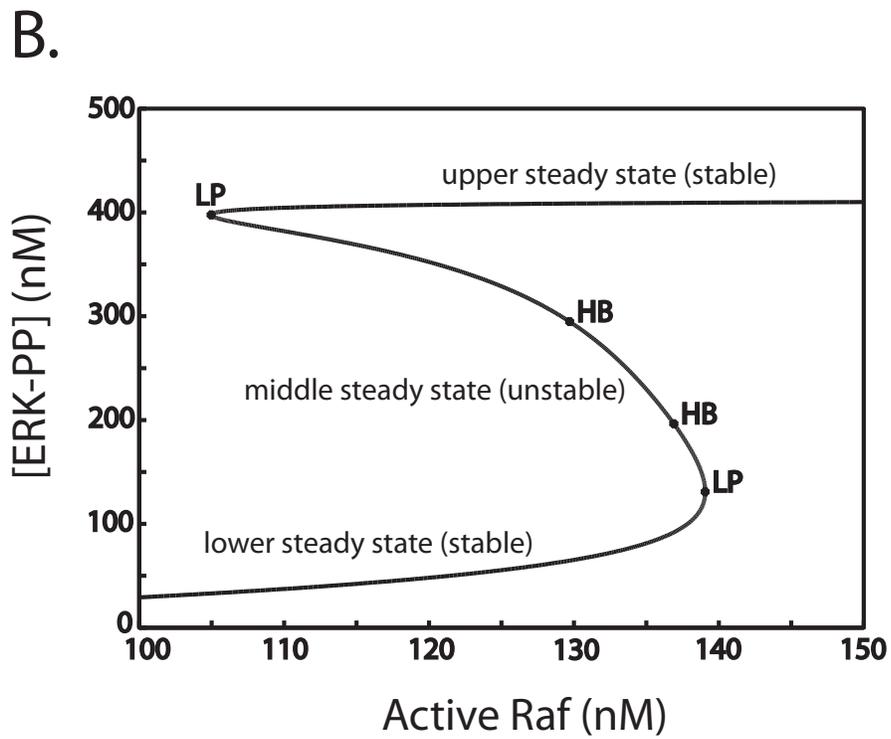

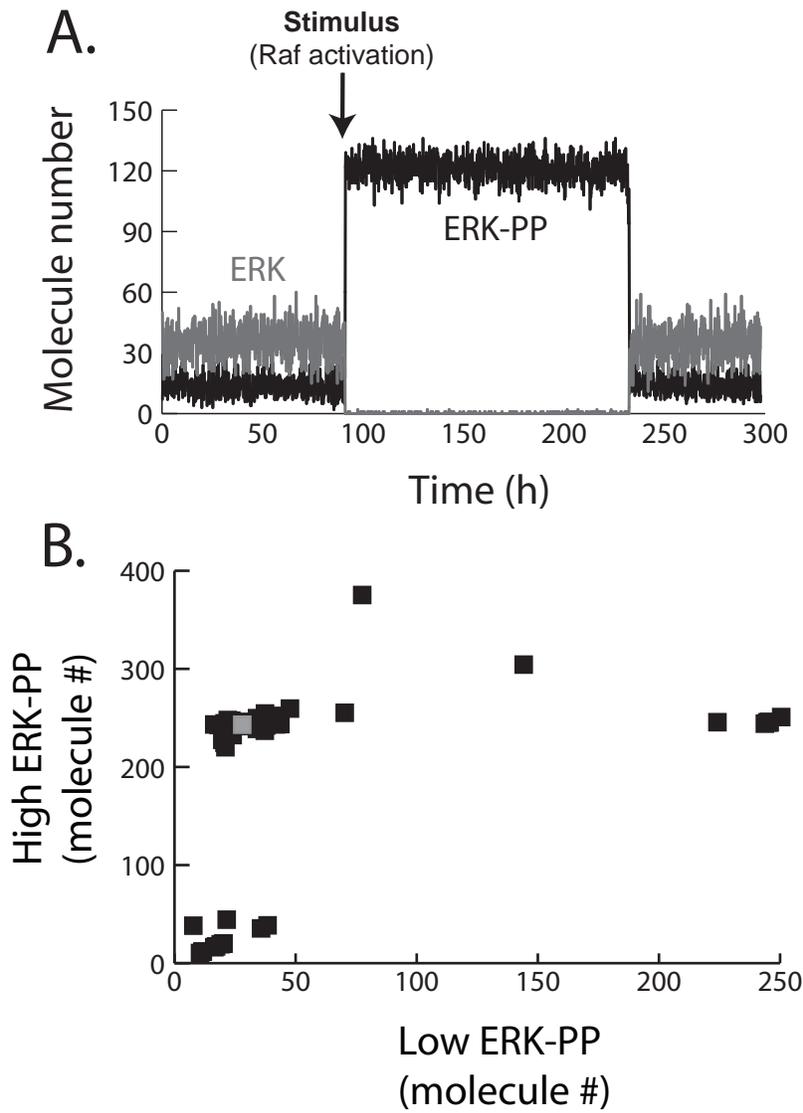

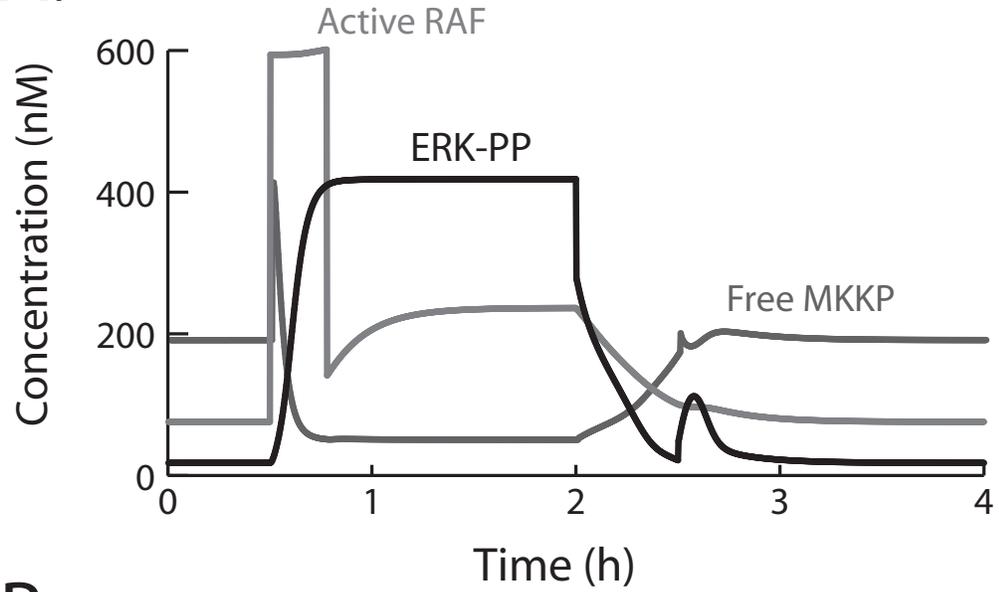

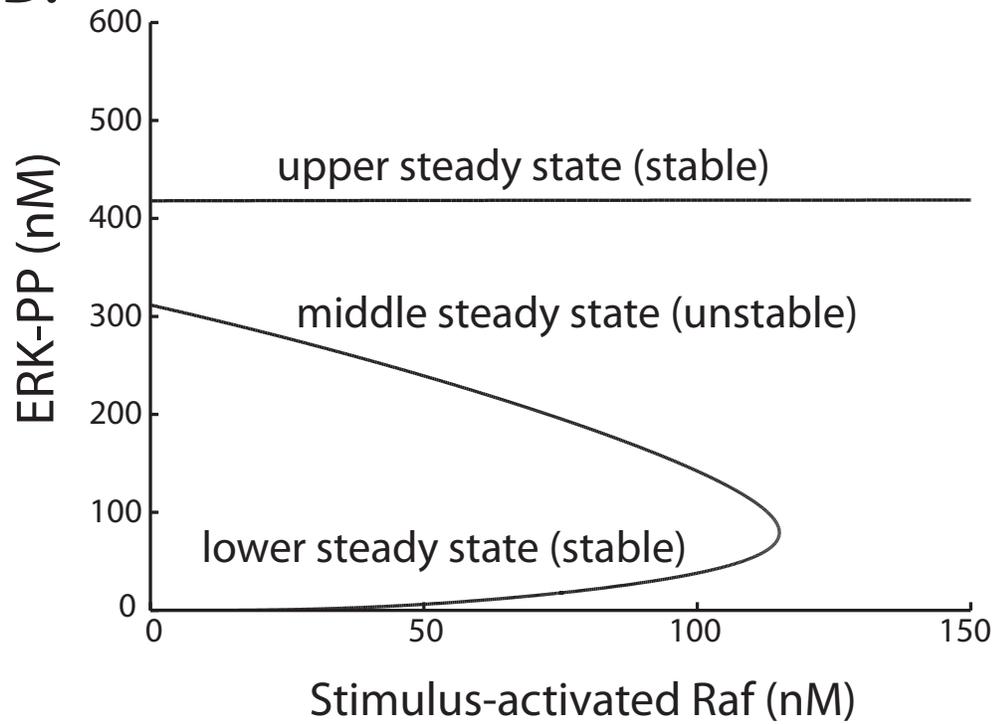

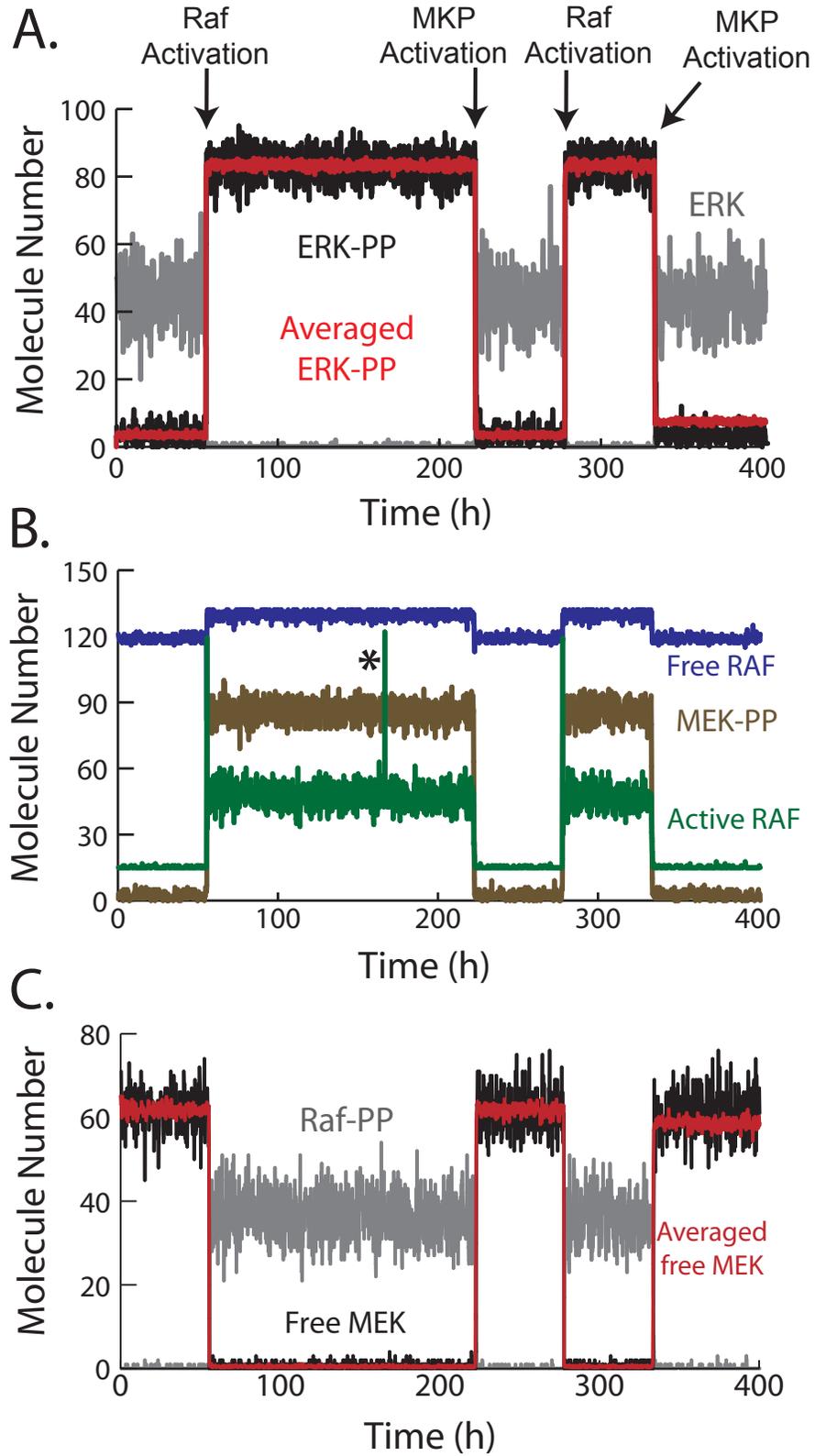

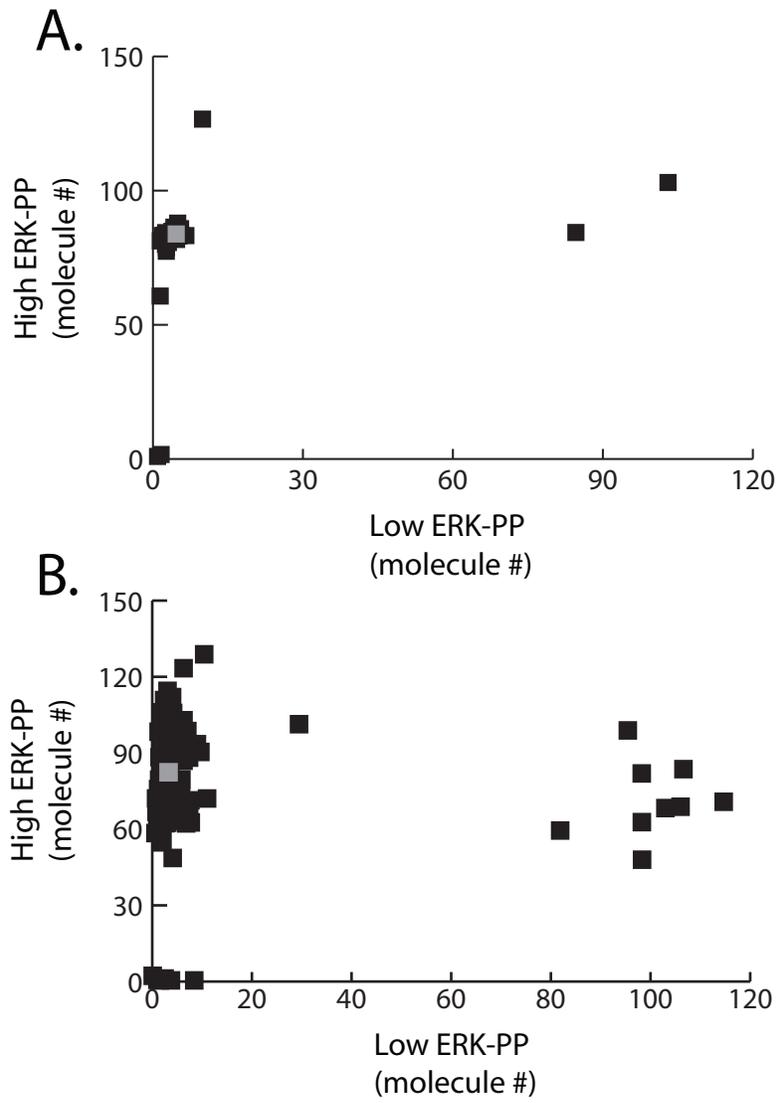

A.

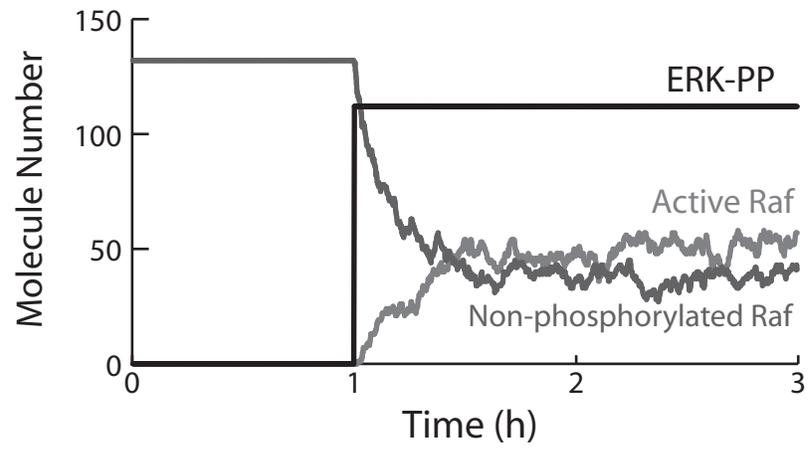

B.

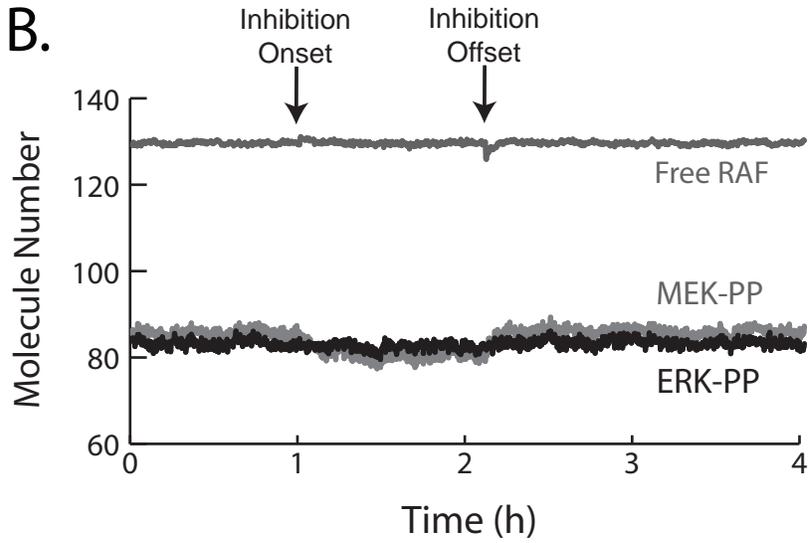